\documentclass{ifacconf}

\usepackage{graphicx}      
\usepackage{natbib}        

\usepackage{amsmath} 
\usepackage{amssymb}  

\usepackage{romannum}
\usepackage{algorithm} 
\usepackage{algpseudocode} 
\usepackage{arydshln}
\usepackage{amsfonts}
\DeclareMathOperator*{\argmin}{arg\,min}

\usepackage{lipsum}
\usepackage{color}
\begin{document}
\begin{frontmatter}

\title{Scheduling Dosage of Proton Pump Inhibitors
Using Constrained Optimization With Gastric Acid Secretion Model\thanksref{footnoteinfo}}

\thanks[footnoteinfo]{The last author acknowledges support by the National Science Foundation grant number CMMI-1904394.}

\author[First]{Yutong Li} 
\author[Second]{Nan Li} 
\author[Third]{Anouck Girard}
\author[Fourth]{Ilya Kolmanovsky}

\address[First]{Department of Aerospace Engineering, 
   University of Michigan, Ann Arbor, MI 48109, USA (e-mail: wilson420813@gmail.com)}
\address[Second]{Department of Aerospace Engineering, 
    Auburn University, Auburn, AL 36849, USA (e-mail: nanli@auburn.edu)}
    \address[Third]{Department of Robotics, 
   University of Michigan, Ann Arbor, MI 48109, USA (e-mail: anouck@umich.edu)}
\address[Fourth]{Department of Aerospace Engineering, 
   University of Michigan, Ann Arbor, MI 48109, USA (e-mail: ilya@umich.edu)}

\begin{abstract}                
Dosage schedule of the Proton Pump Inhibitors (PPIs) is critical for gastric acid disorder treatment. In this paper, we develop a constrained optimization based approach for scheduling the PPIs dosage. In particular, we exploit a mathematical prediction model describing the gastric acid secretion, and use it within the optimization algorithm to predict the acid level. The dosage of the PPIs which is used to enforce acid level constraints is computed by solving a constrained optimization problem. Simulation results show that the proposed approach can successfully suppress the gastric acid level with less PPIs intake compared with the conventional fixed PPIs dosage regimen, which may reduce the long-term side effects of the PPIs.

\end{abstract}


\end{frontmatter}

\section{Introduction}

Over the last two decades, drug development and treatment of gastric acid related illnesses have become of interest as twenty million people suffer from the gastric acid disorder in the US, and 770,000 death cases of gastric cancer are reported globally in 2020 \citep{morgan2022current}. One class of drugs known as Proton Pump Inhibitors (PPIs) has been proven to be highly effective in treating gastric acid secretion disorders by raising stomach pH value \cite{reimer2013safety,huang2001pharmacological}. 

Although PPIs have been successfully used for gastric acid disorder treatment, how to optimize their dosage schedule is still a open question. This is important as overdose of the PPIs may cause long-term side effects on the gastric health. One route to design the PPIs dosage is through clinical trials \citep{shin2013pharmacokinetics,lundell2015systematic,shin2008pharmacology}. Another route is to use mathematical modeling to describe and analyze the gastric acid secretion process, and design the dosage schedule based on these prediction models \cite{de1993gastric,livcko1992dual,joseph2003model,sud2004predicting}.

In this paper, we adopt the second route and develop a constrained optimization approach to enforce the gastric acid constraints. By exploiting the gastric acid secretion prediction model and online optimization techniques, a dynamic PPIs dosage schedule will be generated given the patient's gastric state. This is different from the approach used in \cite{sud2004predicting}, where the PPIs dosage schedule is fixed. A potential benefit of our approach is that our dynamic PPIs dosage schedule can reduce the total PPIs intake, which may alleviate their long-term side effect on gastric health. On the other hand, our approach is able to personalize the PPIs dosage schedule, which is achieved by adapting the prediction model parameters according to individual patient's physical condition and disease symptoms.

The main contributions of this paper include: 1) establishing a constrained optimization based PPIs dosage scheduling approach, 2) comparing the effectiveness of the proposed approach with a fixed PPI dosage based approach, 3) demonstrating the capability of the proposed approach for personalized PPIs dosage scheduling.  

\section{The gastric acid secretion model}\label{sec:2}
\begin{figure}[thpb]
      \centering
      \includegraphics[width=250 pt]{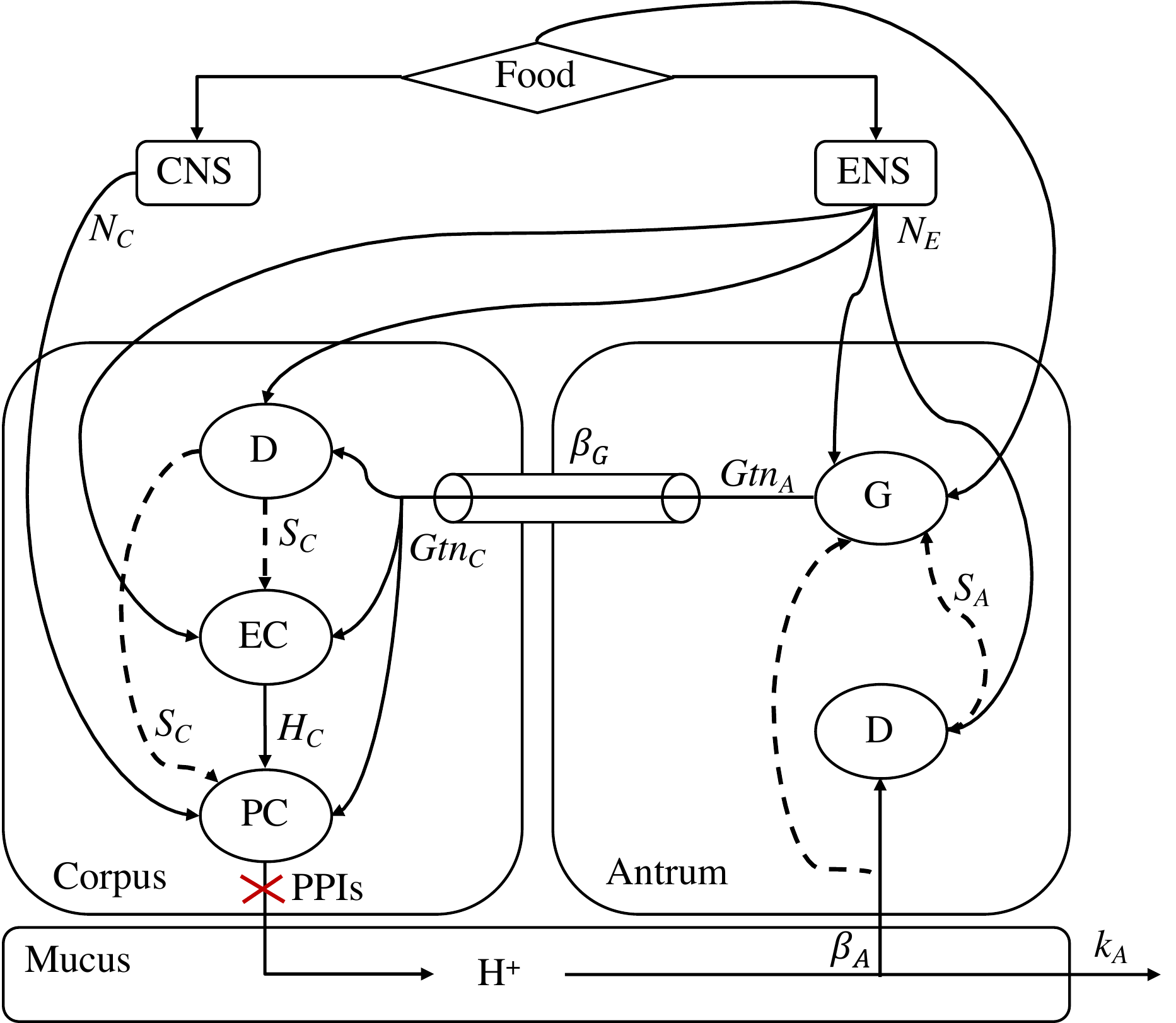}
      \caption{Diagram of regulation of gastric acid secretion. G cells in the antrum secrete gastrin ($Gtn_A$). Gastrin stimulates enterochromaffin-like cells (EC) to release histamine ($H_C$). The gastric acid ($H^{+}$) is secreted from the parietal cells (PC). Somatostatin ($S_A$ and $S_C$) acts as the acid secretion inhibitor. $\beta$ represents the transport rate; $k_A$ is the washout rate of acid. Central and Enteric Neural Stimuli (CNS and ENS, respectively) are stimulated by the food intake. Solid and dashed arrows represent positive and negative stimuli, respectively. Red cross represents the location of the PPIs action.}
      \label{fig:GastricMdl}
\end{figure}

We adopt a dynamic model of gastric acid secretion with PPIs inputs from \cite{sud2004predicting}, \cite{joseph2003model}. The gastric acid secretion process is illustrated in Fig. \ref{fig:GastricMdl}.  There are two main regions represented in the stomach model, namely corpus and antrum. After taking food, the activity levels of both Central Neural Stimuli (CNS) and Enteric Neural Stimuli (ENS) increase, which serve as the stimuli for gastric acid secretion. In the antrum, G cells are first stimulated by ENS to secrete gastrin. Gastrin is released to both the antral blood capillaries and the corpus. The gastrin released to the corpus will stimulate the parietal cells to secrete gastric acid. Enterochromaffin-like (ECL) cells are stimulated by both the gastrin and ENS, and release histamine. Histamine enhances acid secretion with gastrin in a paracrine manner,
and also increases the stimulation intensity of gastrin for the parietal cells. To balance the acid secretion and maintain the gastric acid at a normal level, D cells in the corpus release the somatostatin, which is the inhibitor of the acid secretion. As shown in Fig. \ref{fig:GastricMdl}, PPIs inhibit the proton pumps in the membrane of parietal cells, and thus suppress the acid release.

\subsection{Hormonal effectors dynamics}
The model adopted in this paper is based on Michaelis-Menten kinetics to represent effector secretion in response to
stimuli \citep{joseph2003model,sud2004predicting,johnson2011original}. In this model, the effector secretion is dependent on the corresponding stimuli in a dose-dependent manner. In general, this stimulated secretion can be represented by the term of the form $\frac{K_S[E]}{[E]+\alpha_S}$, where $K_S$ is the maximal rate of secretion of $S$ due to stimulation
with $E$, $[E]$ is stimulator concentration and $\alpha_S$ is the Michaelis constant of the effector, which equals to effector concentration level at which stimulator secretion rate is half maximal. On the other hand, the inhibitory dynamics of effector secretion is captured by the term of the general form $\frac{1}{1+\frac{[I]}{k_I}}$, where $[I]$ is inhibitor concentration and $k_I$ is the dissociation constant of $[I]$. 

The model also represents the loss of effectors via transport and/or degradation as directly proportional to
the effectors concentration.
The dynamics of effectors are defined by using the following equations. 

For antral gastrin ($[Gtn_A(t)]$):
{\small\begin{align} 
\begin{split}\label{equ:AntralGastrin}
    \frac{d[Gtn_A(t)]}{dt}=&\frac{N_GK_{NG_1}[N_E(t)]}{([N_E(t)]+\alpha_{NG_1})(1+\frac{[S_A(t)]}{k_{SG}})(1+\frac{[A_C(t)]^2}{[A_C(t)]^2+k_{AG}^2})} \\&+\frac{N_GK_{NG_2}[N_C(t)]}{([N_C(t)]+\alpha_{NG_2})(1+\frac{[S_A(t)]}{k_{SG}})(1+\frac{[A_C(t)]^2}{[A_C(t)]^2+k_{AG}^2})} \\&+\frac{N_GK_{FG}Fd(t)}{(Fd(t)+\alpha_{FD})(1+\frac{[S_A(t)]}{k_{SG}})(1+\frac{[A_C(t)]^2}{{[A_C(t)]}^2+k_{AG}^2})} \\&-(k_G+\beta_G)[Gtn_A(t)].
\end{split}
\end{align}}

For corpal gastrin ($[Gtn_C(t)]$):
{\small\begin{align} 
\begin{split}\label{equ:CorpalGastrin}
    \frac{d[Gtn_C(t)]}{dt}=\beta_G[Gtn_A(t)]-k_G[Gtn_C(t)].
\end{split}
\end{align}}

For antral somatostatin ($[S_A(t)]$):
{\small\begin{align} 
\begin{split}\label{equ:AntralSomatostatin}
    \frac{d[S_A(t)]}{dt}=&\frac{N_{DA}K_{AS}[A_A(t)]}{([A_A(t)]+\alpha_{AS})(1+\frac{[S_A(t)]}{k_{SS}})(1+\frac{[N_C(t)]}{k_{NS}})} \\&+\frac{N_{DA}K_{NS_1}[N_E(t)]}{([N_E(t)]+\alpha_{NS_1})(1+\frac{[S_A(t)]}{k_{SS}})(1+\frac{[N_C(t)]}{k_{NS}})} \\&-k_S[S_A(t)].
\end{split}
\end{align}}

For corpal somatostatin ($[S_C(t)]$):
{\small\begin{align} 
\begin{split}\label{equ:CorpalSomatostatin}
    \frac{d[S_C(t)]}{dt}=&\frac{N_{DC}K_{GS}[Gtn_C(t)]}{([Gtn_C(t)]+\alpha_{GS})(1+\frac{[S_C(t)]}{k_{SS}})(1+\frac{[N_C(t)]}{k_{NS}})}\\&+\frac{N_{DC}K_{NS_2}[N_E(t)]}{([N_E(t)]+\alpha_{NS_2})(1+\frac{[S_C(t)]}{k_{SS}})(1+\frac{[N_C(t)]}{k_{NS}})}\\&-k_S[S_C(t)].
\end{split}
\end{align}}

For histamine ($[H_C(t)]$):
{\small\begin{align} 
\begin{split}\label{equ:Histamine}
    &\frac{d[H_C(t)]}{dt}=\\&\frac{N_{E}K_{NH}[N_E(t)]}{([N_E(t)]+\alpha_{NH})(1+\frac{[S_C(t)]}{k_{SH}})}+\frac{N_{E}K_{GH}[Gtn_C(t)]}{([Gtn_C(t)]+\alpha_{GH})(1+\frac{[S_C(t)]}{k_{SH}})}\\&-k_H[H_C(t)].
\end{split}
\end{align}}

Here $N_G$ is the number of G cells, $N_E$ is the number of ECL cells, $N_c$ is the number of CNS cells, $K_{NG_1}$ is the maximal secretion rate of gastrin per cell due to the stimulation by ENS, $\alpha_{NG_1}$ and $\alpha_{NG_2}$ are gastrin Michiaelis constants of ECL and parietal cells, respectively, $k_{SG}$ is the dissociation constant of the somatostatin, $k_G$ is the degradation rate of the gastrin, and $\beta_G$ is the transport rate of the gastrin from anturm to corpus. 

\subsection{Acid and bicarbonate dynamics}
The gastric acid is secreted from the parietal cells which are stimulated by gastrin, histamine, and central neural stimuli. The somatostatin acts as the inhibitor of the acid secretion. The gastric acid diffusion from the corpus region to the antral region occurs at a constant rate $\beta_A$. Meanwhile, 
bicarbonate buffers the level of the acid leading to further loss of acid, which is represented by a mass action terms, $hb[A_C(t)][B_C(t)]$ and $hb[A_A(t)][B_A(t)]$.
In addition, the following
term of $\frac{H_C(t)}{H_C(t)+\alpha_H}$ is used to describe the potentiation of histamine on gastrin-mediated gastric acid secretion. The rate of change of gastric acid in the corpus and antrum are described as follows.

For corpal acid ($[A_C(t)]$):
{\small\begin{align} 
\begin{split}\label{equ:CorpalAcid}
    &\frac{d[A_C(t)]}{dt}=\\&\left(\frac{PP_n(t)\times N_{P}K_{HA}[H_C(t)]}{([H_C(t)]+\alpha_{HA})(1+\frac{[S_C(t)]}{k_{SA}})}+\frac{PP_n(t)\times N_{P}K_{NA}[N_C(t)]}{([N_C(t)]+\alpha_{NA})(1+\frac{[S_C(t)]}{k_{SA}})}\right)\\&+\times\left(\frac{[H_C(t)]}{[H_C(t)]+\alpha_H}\right)\left(\frac{PP_n(t)\times N_{P}K_{GA}[Gtn_C(t)]}{([Gtn_C(t)]+\alpha_{GA})(1+\frac{S_C(t)}{k_{SA}})}\right)\\&-hb[A_C(t)][B_C(t)]-\beta_A[A_C(t)].
\end{split}
\end{align}}

For antral acid ($[A_A(t)]$):
{\small\begin{align} 
\begin{split}\label{equ:AntralAcid}
    \frac{d[A_A(t)]}{dt}=\beta_A[A_C(t)]-k_A[A_A(t)].
\end{split}
\end{align}
}

Here $N_P$ is the number of parietal cells, $K_{HA}$ is the maximal secretion rate of acid per cell due to the histamine, $\alpha_{HA}$ and $\alpha_{NA}$ are histamine and acid Michiaelis constants, respectively, $k_{SA}$ is the dissociation constant of the somatostatin from receptors on parietal cells, $\beta_A$ is the transport rate of the acid from corpus to anturm, and $k_A$ is the wash out rate of the gastrin acid. 

Note the $PP_n(t)\in [0,1]$ is the factor that indicates the fraction
of pumps in the parietal cells that are still active to secret acid into
the stomach lumen. The dynamics of active proton pump cycling during treatment is given by
{\small\begin{align} 
\begin{split}\label{equ:ProtonPump}
    \frac{d(PP_n(t))}{dt}=&K_{deg}-K_r\times PPI(t)\times PP_n(t)-K_{deg}\times PP_n(t),
\end{split}
\end{align}}

where  $K_{deg}$
is the decay rate of proton pump, and $K_r$ is the bimolecular rate constant of PPIs. $PPI(t)$ represents the PPIs blood concentration depending on PPIs dosing regimen, which is to be designed in the next section.

Bicarbonate secretion, which is stimulated by the CNS, is represented as follows.

For corpus bicarbonate ($[B_C(t)]$):
{\small\begin{align} 
\begin{split}\label{equ:CorpusBicarbonate}
    \frac{d[B_C(t)]}{dt}=\frac{k_{bc}[N_C(t)]}{N_C(t)+\alpha_{NB}}-hb[A_C(t)][B_C(t)]-k_B[B_C(t)].
\end{split}
\end{align}}

For antral bicarbonate ($[B_A(t)]$):
{\small\begin{align} 
\begin{split}\label{equ:CorpusBicarbonate}
    \frac{d[B_A(t)]}{dt}=\frac{k_{ba}[N_C(t)]}{N_C(t)+\alpha_{NB}}-hb[A_A(t)][B_A(t)]-k_B[B_A(t)].
\end{split}
\end{align}}

Here $\alpha_{NB}$ is the bicarbonate Michiaelis constant and $k_B$ is the wash out rate of the bicarbonate. 

\subsection{Central and enteric neural stimuli}
The activities of central and enteric neural stimuli, $[N_C(t)]$ and $[N_E(t)]$,
which are evoked by food intake $Fd(t)$, are expressed as follows. 

For central neural stimuli ($[N_C(t)]$):
{\small
\begin{align} 
\begin{split}\label{equ:CNS}
    \frac{d[N_C(t)]}{dt}=&\left(\frac{N_{1}[Fd(t)]}{(Fd(t)+k_1^{Fd})(1+\frac{A_C(t)^2}{A_C(t)^2+k_{AN1}^2})}\right)\\&-k_{N_C}[N_C(t)]+Bas_1.
\end{split}
\end{align}}

For enteric neural stimuli ($[N_E(t)]$):
{\small
\begin{align} 
\begin{split}\label{equ:ENS}
    \frac{d[N_E(t)]}{dt}=&\left(\frac{N_{2}[Fd(t)]}{(Fd(t)+k_2^{Fd})(1+\frac{A_C(t)^2}{A_C(t)^2+k_{AN2}^2})}\right)\\&-k_{N_E}[N_E(t)]+Bas_2.
\end{split}
\end{align}}

Here $Fd$ is the food intake profile, $k_1^{Fd}$ and $k_2^{Fd}$ are the CNS and ENS Michiaelis constant with respect to the food intake, and $Bas_1$ and $Bas_2$ are the basal neural activity constant of the CNS and ENS, respectively. 

\section{Constrained optimization Based PPIs Dosage scheduling}\label{sec:3}
In this section, we develop a PPIs dosage scheduling approach based on constrained optimization techniques exploiting the gastric acid secretion model, to enforce gastric acid constraint with minimal PPIs intakes.  

We adopt a recommended PPIs twice-daily dosing regimen (drug is administered twice a day at fixed
times throughout the day) \citep{sud2004predicting}. The following equation is used to account for the accumulation of drug in the
blood following PPIs administration:
{\small\begin{align} 
\begin{split}\label{equ:PPI_concentration}
    PPI(t) = \sum_{i=1}^{n_d} \sum_{j=1}^2 \frac{D_{i,j}}{V\times m}e^{-K_{el}(t-t_{i,j})},  
\end{split}
\end{align}}

where $D_{i,j}$ and $t_{i,j}$ are the dosage and dosing time at the $i^{\text{th}}$ day's $j^{\text{th}}$ administration (in our twice-daily dosing regimen, $j\in\{1,2\}$), respectively, $m$ is the molecular weight of the PPIs, $V$ is the volume of distribution, and $K_{el}$ is the elimination constant, $t$ is the time since the first dosage is administrated. 

Our objective is to enforce a specified the constraint on corpal acid levels by using minimal PPIs intakes as this reduces the risk of long-term side effects caused by the PPIs overdose. To achieve this, we formulate the following optimization problem to online compute the PPIs dosage $D_{i,j}$ at each dosing time $t_{i,j}$,
{\small\begin{align}\label{equ:ConstrOpt}
   & D^*_{i,j}  = \argmin_{D_{i,j} \in \mathcal{D}} \ D_{i,j}\\
     \text{s.t. }\, &\frac{dx(t)}{dt}=f(x(t),F_d(t),PPI(t,D_{i,j})),\ \forall t\in [t_{i,j},t_{i,j}+T_p], \nonumber \\ & A_C(t) \leq 0.035\ [\text{M}],\ \forall t\in [t_{i,j},t_{i,j}+T_p]\nonumber.
\end{align}}

The system state vector is defined as $x=[Gtn_A,Gtn_C,S_A,\\S_C,H_C,A_C,A_A,B_C,B_A,N_C,N_E]^T$, and $f(\cdot,\cdot,\cdot)$ is the compact form of equations \eqref{equ:AntralGastrin} - \eqref{equ:PPI_concentration}. The set $\mathcal{D}=[0,d^\text{max}]$, where $d^\text{max}$ is the maximum dosage for one-time PPIs intake. $T_p$ is the prediction horizon, which is chosen as the duration between current and next dosing time. For simplicity, we assume $T_p=t_{i,2}-t_{i,1}=t_{i+1,2}-t_{i,2}=12$ hours. $F_d$ is the food intake, which is modeled to represent a typical daily three meal profile as follows \citep{joseph2003model}:
{\small\begin{align} 
\begin{split}\label{equ:Fd}
    Fd(t)=& 1.6(1 + \tanh(\pi[t-(f_\text{fl}+19)]))e^{-0.5(1+3.5[t-(f_\text{fl}+19)])} + \\&(1 + \tanh(\pi[t-(f_\text{fl}+13)]))e^{-0.5(1+3.5[t-(f_\text{fl}+13)])} + \\&0.4(1 + \tanh(\pi[t-(f_\text{fl}+7)]))e^{-0.5(1+3.5[t-(f_\text{fl}+7)])},
\end{split}
\end{align}}

where $f_\text{fl} = 24\text{floor}(\frac{t}{24})$.

Note that \eqref{equ:ConstrOpt} can be extended to address more general problems, e.g., simultaneously optimize the PPIs dosage, dosing time and food intake profile. These are key effectors of the treatment plan, see e.g., \cite{freedberg2017risks}. The optimization problem formulated in \eqref{equ:ConstrOpt} can also include data-driven gastric acid secretion prediction models. Within this paper, we aim at demonstrating the feasibility of using constrained optimization with mathematical prediction model for scheduling PPIs dosage. We restrict our problem formulation as in \eqref{equ:ConstrOpt} and leave these extensions for the future work.

{
\begin{algorithm}
        \caption{Bisection method to solve \eqref{equ:ConstrOpt}}
         \textbf{Input:} {$\ x(t).\ PPI(t)$}\\
        \textbf{Output:}{$\ D_{i,j}(t)$}
        \label{algorithm:bisection}
        \begin{algorithmic}[1] 
            \If {$t==t_{i,j}$}
            \State $D_{i,j}^\text{tempt} \gets 0,\ \Bar{\lambda}\gets d^{\text{max}},\ \underline{\lambda}\gets 0,\ CV \gets 1$
            \While {$(\Bar{\lambda}-\underline{\lambda}>\delta)\ ||\ (CV == 1) $}
            \State $\text{Run simulations over prediction horizon}\ T_p \text{ with } \newline
             \hspace*{3em}D_{i,j}^\text{tempt}$ \text{using} \eqref{equ:AntralGastrin} \text{-}\eqref{equ:PPI_concentration}\text{ with food intake }\eqref{equ:Fd}
            \If {$\text{Constraint violation happens}$}
                \State $\underline{\lambda}\gets D_{i,j}^\text{tempt}$
            \Else
                \State $\Bar{\lambda}\gets D_{i,j}^\text{tempt}$
                \State $CV\gets 0$
            \EndIf
            \State $D_{i,j}^\text{tempt}\gets \frac{1}{2}(\Bar{\lambda}+\underline{\lambda})$
            \EndWhile
            \State $D_{i,j}(t) \gets D_{i,j}^\text{tempt}$
        \EndIf
        \end{algorithmic}
\end{algorithm}
}

Algorithm \ref{algorithm:bisection} aims to find a feasible value $D_{i,j}(t)$ between $0$ and $d^\text{max}$ that is as close to
$0$ as possible through a bisection method. 
By administrating $d^\text{max}$ at each dosing time $t_{i,j}$, we have $A_C(t) \leq 0.035\ [\text{M}],\ \forall t\in[t_{i,j},t_{i,j}+T_p+k]$, where $k>0$. Taking a large dosage indeed leads to more suppression of the acid secretion, however, may increase the risk of PPIs long term side effects. Algorithm \ref{algorithm:bisection} aims to find the minimum volume of PPIs which is sufficient to enforce the acid level below the prescribed value. 

We note that the parameters in the gastric acid secretion model could be estimated offline based on individual patient historical data, see e.g., \cite{sud2004predicting,joseph2003model}. Then the proposed approach could be used for personalizing the treatment plan. 

We also note that in practice, patient's feedback can be included in the loop to dynamically adjust constraints to be enforced. For instance, after a certain treatment period, the symptoms alleviation extent can be evaluated by the patient. If the patient does not feel relief, we decrease the constraint value for $A_c$ in \ref{equ:ConstrOpt} until the patient feels better. By using the patient's feedback in the loop, we could implement the proposed approach without additional measurements. We also note that the algorithm could be executed for several food intake scenarios \eqref{equ:Fd}, model parameter values and initial conditions $x(t)$ with largest $D_{i,j}(t)$ chosen as the solution. Details and scalability of such an approach are left to be addressed in the future work. 

\section{Simulation results}\label{sec:4}
In this section, we report simulation results from applying the proposed constrained optimization approach to PPIs dosage scheduling to enforce gastric acid constraints in the simulation. We use the gastric acid secretion model in Section \ref{sec:2} as the plant model. The C++ Runge-Kutta solver odeint described in the paper by \cite{ahnert2011odeint} is leveraged for integrating the ordinary
differential equations \eqref{equ:AntralGastrin}-\eqref{equ:ENS}. 

\begin{figure}[thpb]
      \centering
      \includegraphics[width=250pt]{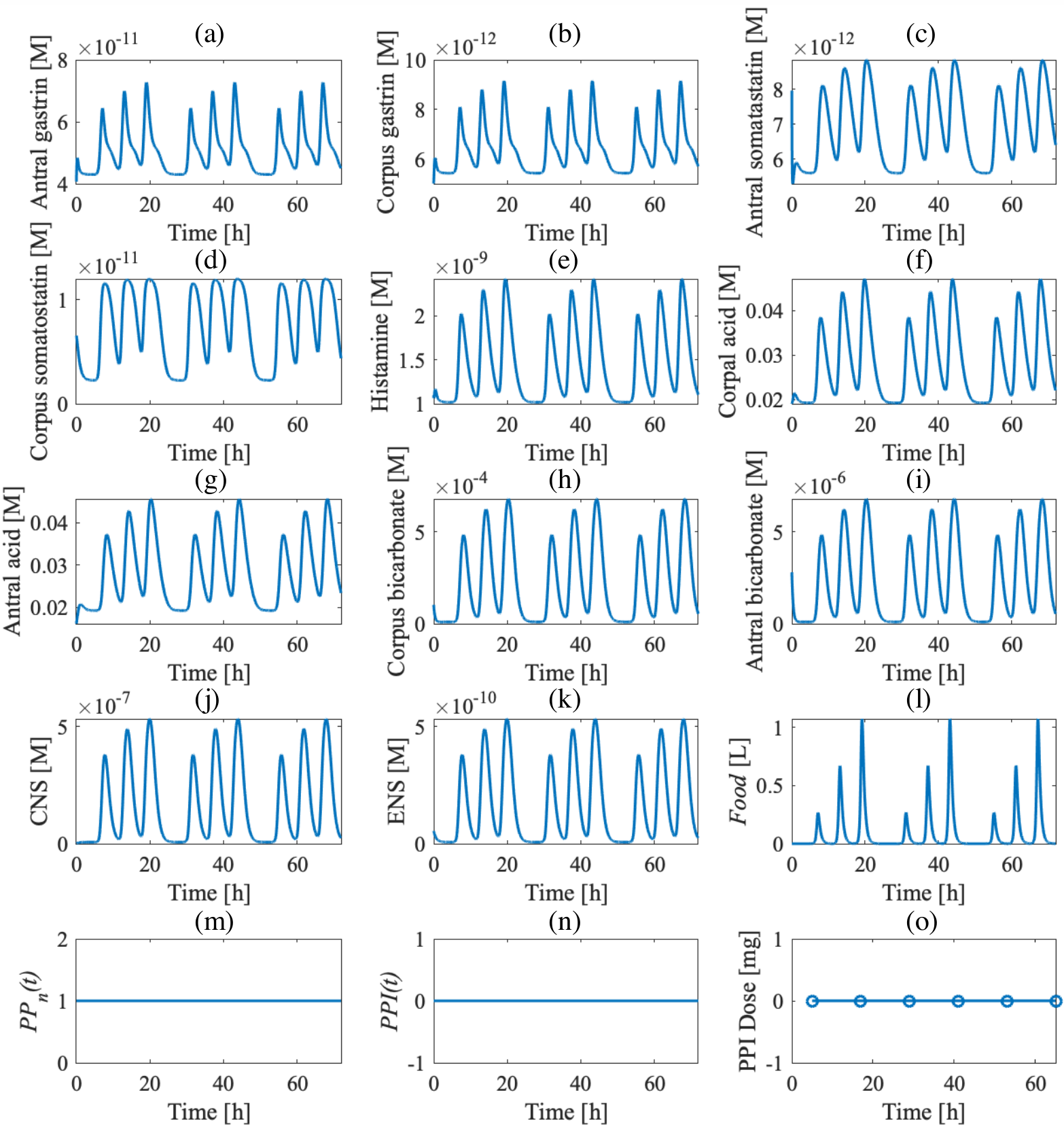}
      \caption{Model validation of the gastric acid secretion model.}
      \label{fig:baseline}
\end{figure}

We first qualitatively evaluate the gastric acid secretion model in Section \ref{sec:2} by administrating no PPIs. This simulation result provides us the baseline condition of patient's gastric activities only with the stimulation of food intakes. Simulations are performed over 3 days, and the results of effectors, acid, bicarbonate and neural stimuli are shown in Fig. \ref{fig:baseline}. 

A food intake in \eqref{equ:Fd} is used to represent a typical daily three meal profile, which is shown in Fig. \ref{fig:baseline}(l) and also used in \cite{sud2004predicting}. Note that the PPIs dosage is set to zero, thus all pumps in a parietal cell are in active mode ($PP_n=1$) and the PPIs concentration is zero ($PPI=0$), as shown in Figs. \ref{fig:baseline}(o), (m) and (n), respectively. We observe that the increases in neural activities as in Figs. \ref{fig:baseline}(j) and (k), are closely correlated with the food intakes. 
Food ingestion together with the increasing neural activities promote the release of the antral gastrin (see \eqref{equ:AntralGastrin}), which is
transported to the corpus (see \eqref{equ:CorpalGastrin}). The model also reproduces a characteristic reciprocal
behavior of gastrin and antral somatostatin
that is observed in in vivo and in vitro systems \citep{zavros2002gastritis}. This highlights the antagonistic relationship
between the two effectors: gastrin release occurs
first, followed by the somatostatin increase, which down-regulates
gastrin secretion. This relationship is modeled in \eqref{equ:AntralSomatostatin} and \eqref{equ:CorpalSomatostatin}, and is verified with the simulation results shown in Figs. \ref{fig:baseline}(c) and (d), where the delay between the release of gastrin and somatostatin can be observed. The simulation results qualitatively reproduce the ones in \cite{sud2004predicting} despite assumptions on the values of several parameters not reported in \cite{sud2004predicting}.

\begin{figure}[thpb]
      \centering
      \includegraphics[width=250pt]{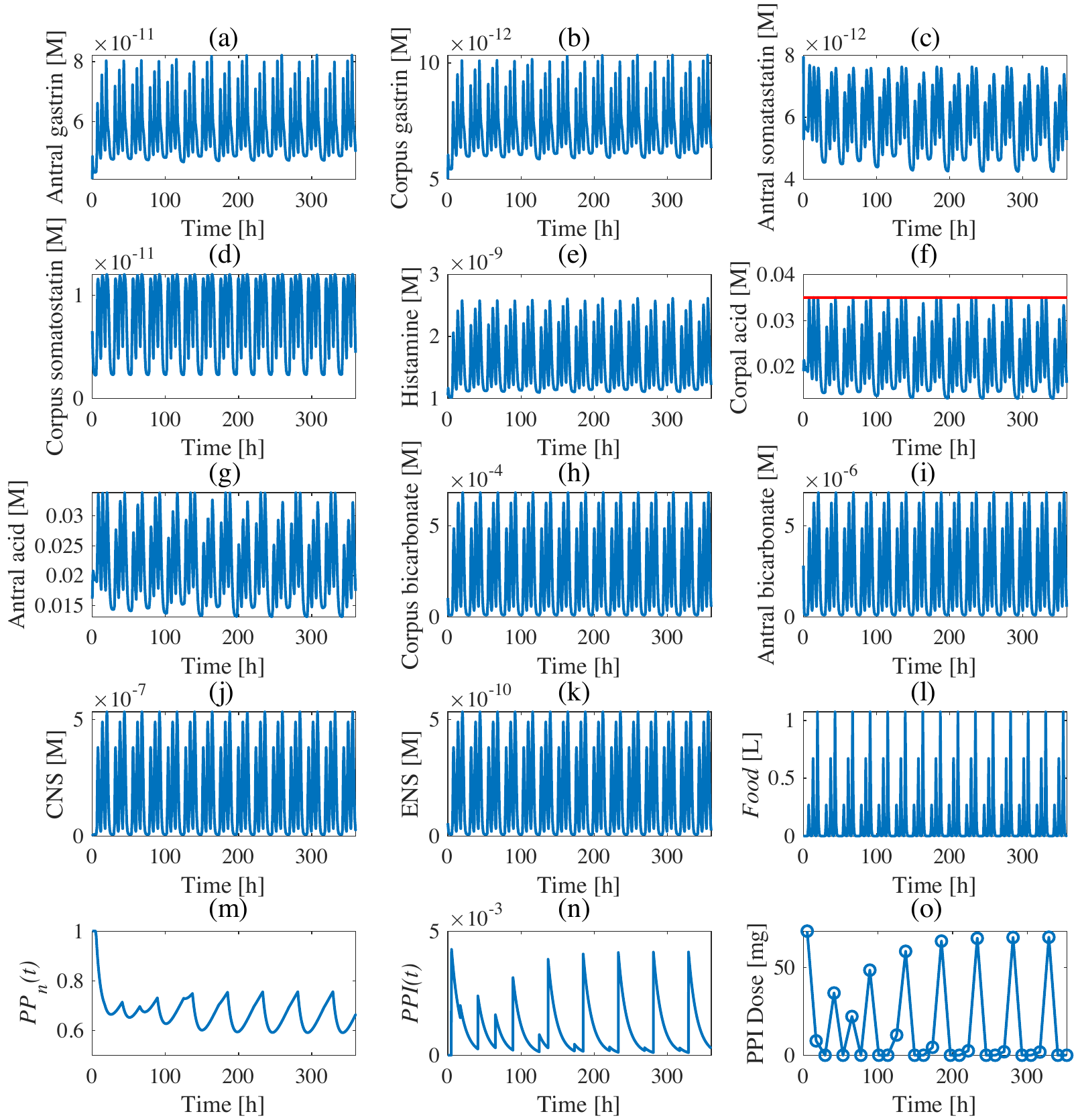}
      \caption{Simulation results of acid secretion suppression via PPIs. The PPIs dosage is calculated via constrained optimization approach in \eqref{equ:ConstrOpt}.}
      \label{fig:Patient35CO}
\end{figure}

Next, we apply the proposed PPIs dosage scheduling approach in Section \ref{sec:3} \eqref{equ:ConstrOpt} to the baseline model to enforce the corpal acid constraint of $A_c(t)\leq 0.035\ \text{[M]}, \forall t\in \mathbb{Z}_{\geq 0}$. Note that without PPIs intervention, the corpal acid reaches 0.048 [M] as shown in Fig. \ref{fig:baseline}(f). In the simulation, we set the daily dosing time at $i^\text{th}$ day as $t_{i,1}=5$ and $t_{i,2}=17$ \citep{sud2004predicting}. The simulation results of a 15 days treatment are shown in Fig. \ref{fig:Patient35CO}. We observe in Fig. \ref{fig:Patient35CO}(f) that the corpal acid is below 0.035 [M] (red solid line) for all time, which means that the constraint in \eqref{equ:ConstrOpt} is enforced. As shown in Fig. \ref{fig:Patient35CO}(o), the dosage of the PPIs is administrated dynamically according to the existing PPIs concentration in the body (in Fig. \ref{fig:Patient35CO}(n)). According to the objective of \eqref{equ:ConstrOpt}, we select the minimal PPIs dosage at each dosing time to enforce the constraint, thus the actual acid will ride on the constraint at some time but never violate it. 

\begin{figure}[thpb]
      \centering
      \includegraphics[width=250pt]{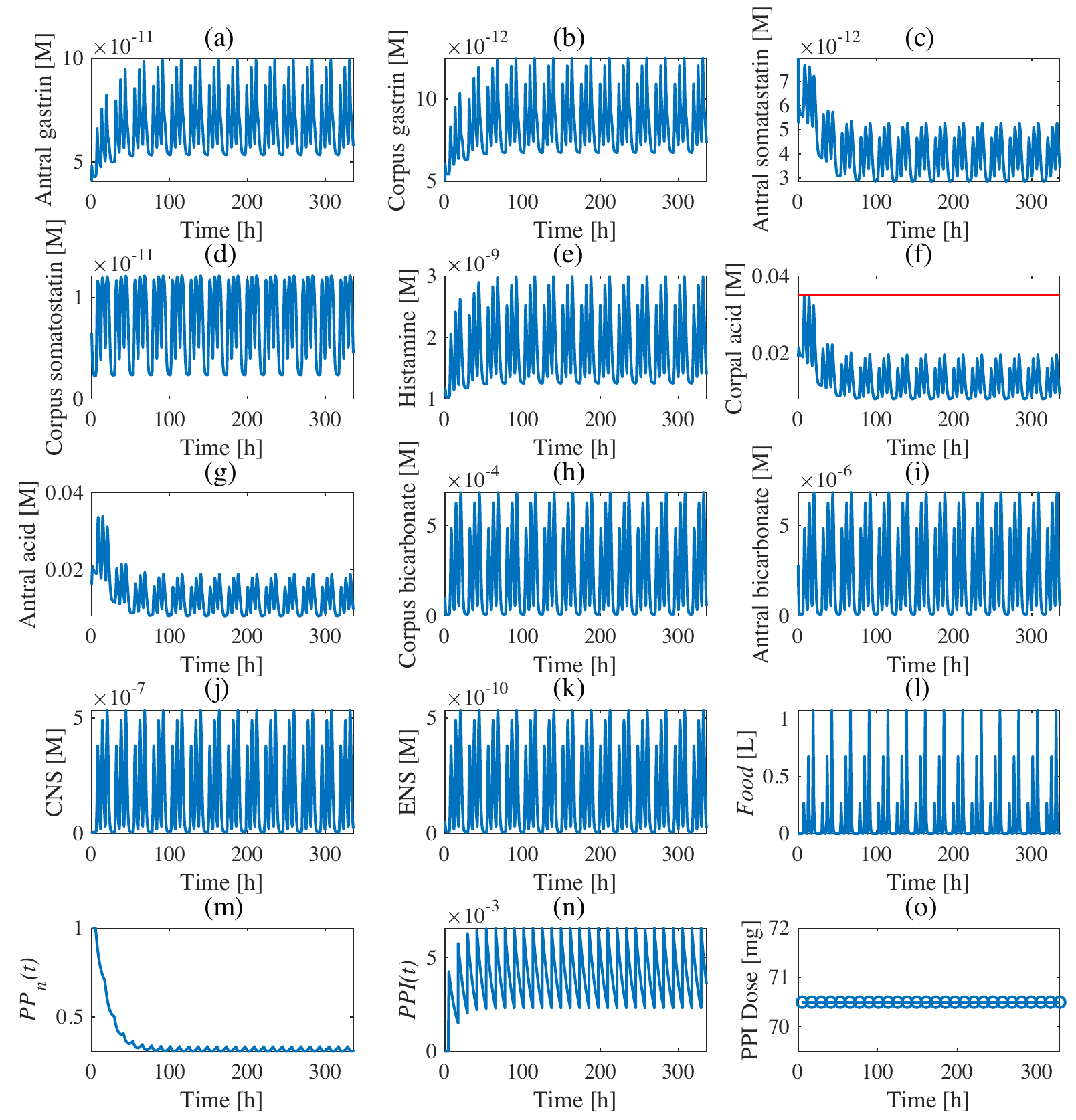}
      \caption{Simulation results of acid secretion suppression via fixed dosage regimen.}
      \label{fig:Patient35Fix}
\end{figure}

\begin{figure}[thpb]
      \centering
      \includegraphics[width=250pt]{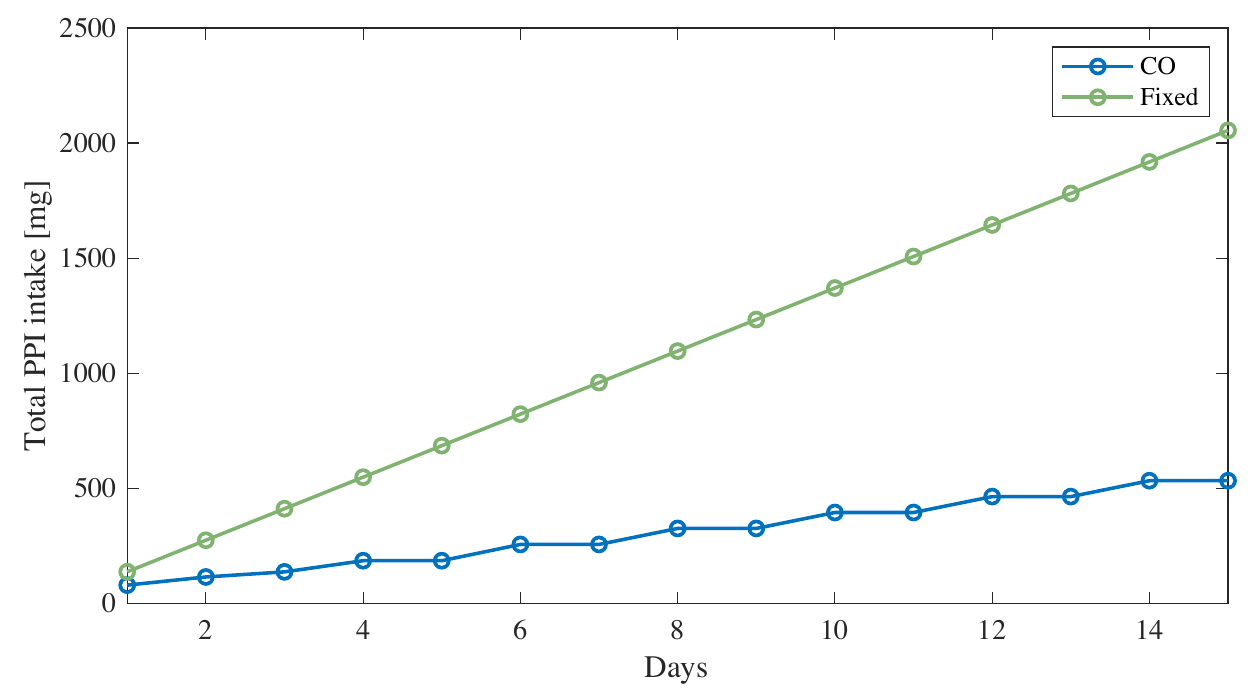}
      \caption{Comparison of total PPIs intake over 15 days with constrained optimization based approach in \eqref{equ:ConstrOpt} and fixed dosage regimen.}
      \label{fig:DosageTotPatient35}
\end{figure}

The simulation results of the dosing regimen with fixed dosage are shown in Fig. \ref{fig:Patient35Fix}. Note that the PPIs dosage is selected as 70.5 mg per dosing time, which is the minimal value to enforce the acid level constraint for all time by using the fixed dosage schedule. Compared with the results from our proposed approach, unnecessary dosage is administrated by the fixed dosage regimen as a large dosage is needed at the beginning of the treatment to establish the PPIs concentration to the level of enforcing the acid constraint. At the same time, this large dosage is unnecessary in the later part of treatment when the existing PPIs concentration is high enough to suppress the acid secretion. The comparison of the total PPIs intake over a 15 days treatment with our constrained optimization based approach and fixed dosage approach is shown in Fig. \ref{fig:DosageTotPatient35}. The proposed approach can reduce PPIs intake compared with the fixed dosage regimen by over 52\% by dynamically scheduling PPIs dosage. This reduced PPIs intake may potentially reduce the patient risk of suffering PPIs long-term side effects.      

\begin{figure}[thpb]
      \centering
      \includegraphics[width=250pt]{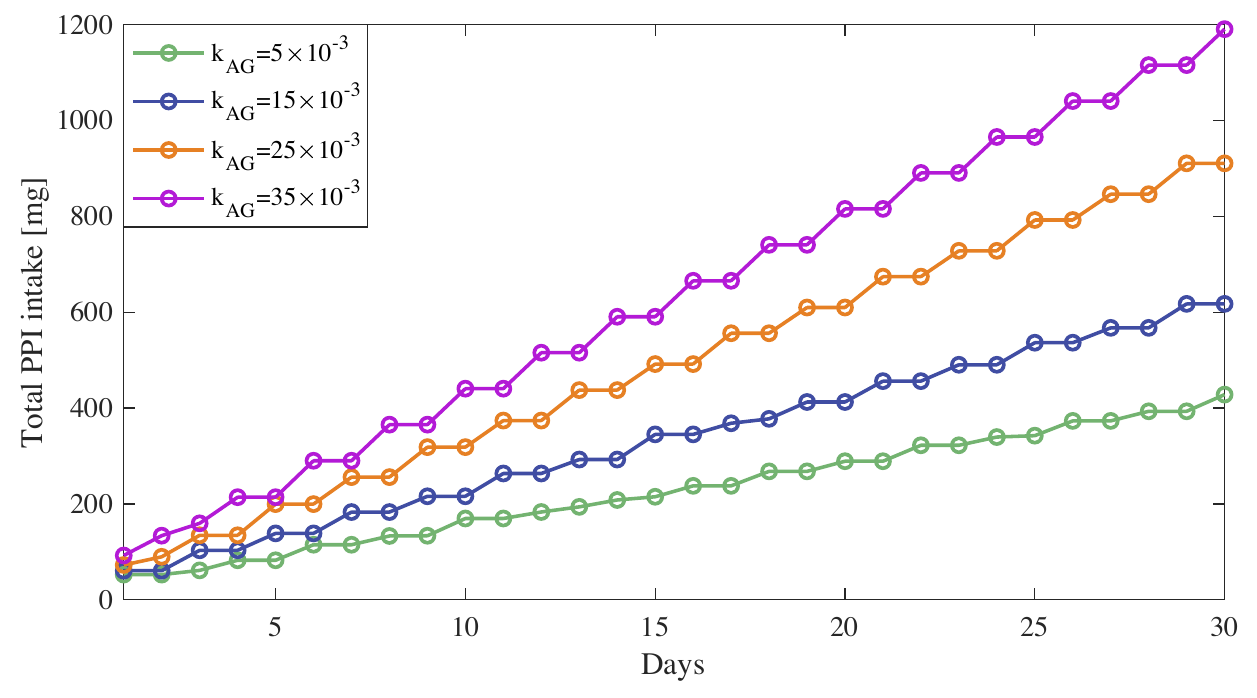}
      \caption{Comparisons of the total PPIs intake of patients with different severity of acid secretion level.}
      \label{fig:DosageTotDiffPatient}
\end{figure}

We next further illustrate the capability of the proposed method to determine personalized dosage regimen based on patient's gastric characteristics, and severity of the illness. Note that $k_{AG}$ in \eqref{equ:AntralGastrin} is the key parameter representing the sensitivity level of gastrin secretion regarding stimulus, e.g., the neural activity and the food intake. Larger values of $k_{AG}$ correspond to higher of severity of the patient's acid over-secretion. Fig. \ref{fig:DosageTotDiffPatient} shows the different dosage regimen obtained by solving \eqref{equ:ConstrOpt} for different patients. 

One of the key observations is that for the more severe patient's acid over-secretion (larger $k_{AG}$), the dosage regimen has more tendency to converge to a periodic one, i.e., the dosing schedule changes from twice daily to every two days as shown in the cases of $k_{AG} = 35\times 10^{-3}$ and $k_{AG} = 25\times 10^{-3}$ in Fig. \ref{fig:DosageTotDiffPatient}. This periodic dosage regimen does not occur in the cases with smaller $k_{AG}$. The reason for this is that for more severe patients, the acid level bounces back rapidly, resulting in a similar acid level (similar initial condition) for \eqref{equ:ConstrOpt} at each time it is used to compute the PPIs dosage. 
 Different from the severe patient cases, for less severe patients, it takes longer time for the acid to bounce back, which causes the optimization to start from different initial conditions at each time PPIs intervention is needed. This leads to an aperiodic dosage solution.
 
This observation provides a useful treatment guidance. Specifically, for the less severe patient, the  dosage regimen needs to be more carefully designed compared to the ones for more severe patients, as the dosage for less severe patient varies day-to-day.

\section{Conclusion}\label{sec:5}
In this paper, we developed a constrained optimization based PPIs dosage scheduling approach to enforce the gastric acid constraint based on the gastric secretion prediction model. We illustrated the acid suppression effectiveness of using the proposed approach, and demonstrated its capability of personalizing acid disorder treatment. Future work will include investigation of the robustness of the proposed approaches, and extending the approach to handle data-driven prediction models.

\bibliography{ifacconnf}
\end{document}